\documentclass[12pt]{iopart}
\usepackage{epsfig}
\usepackage{graphicx}
\usepackage{dcolumn}
\usepackage{color}
\usepackage{multicol}
\usepackage{bm}
\usepackage{iopams}  

\begin{document}

\title{Bound states of Dipolar Bosons in One-dimensional Systems}

\author{A G Volosniev$^1$, J R Armstrong$^{1,2}$, D V Fedorov$^1$, A S Jensen$^1$, M Valiente$^{3}$
and N T Zinner$^1$}
\address{$^1$ Department of Physics and Astronomy - Aarhus University, Ny Munkegade, Building 1520, DK8000 \AA rhus C, Denmark \\
$^2$ Physics Department, Winona State University, 175 W. Mark St Pasteur 120, Winona, MN 55987, USA\\
$^3$ SUPA, Institute of Photonics and Quantum Sciences, Heriot-Watt University, Edinburgh EH14 4AS, United Kingdom}

\begin{abstract}
We consider one-dimensional tubes containing bosonic polar molecules. The long-range dipole-dipole 
interactions act both within a single tube and between different tubes. We consider 
arbitrary values of the
externally aligned dipole moments with respect to the symmetry axis of the tubes. The few-body 
structures in this geometry are determined as function of polarization angles and dipole
strength by using both essentially exact stochastic variational methods and the harmonic approximation. 
The main focus is on the three-, four-, and five-body problems in two or more tubes.
Our results indicate that in the weakly-coupled limit the inter-tube interaction is 
similar to a zero-range term with a suitable rescaled strength. This allows us to 
address the corresponding many-body physics of the system by constructing a 
model where bound chains with one molecule in each tube are the effective degrees
of freedom. This model can be mapped onto one-dimensional
Hamiltonians for which exact solutions are known.
\end{abstract}

\maketitle

\section{Introduction}
Heteronuclear molecules with strong electric dipole moments are a current 
pursuit of the cold atom community
\cite{ospelkaus2008,ni2008,deiglmayr2008,lang2008,ospelkaus2010,ni2010}. 
The dipole-dipole forces in such systems can be externally controlled and 
give access to long-range and anisotropic interactions. Strong interactions
can lead to rapid chemical reaction loss which can, however, be suppressed by considering
low-dimensional trapping potentials \cite{miranda2011,chotia2012}. 
In the case of one or several one-dimensional tubes holding
dipolar particles theoretical works indicate that a number of 
interesting physical states can be realized. Non-trivial Luttinger 
liquid states \cite{citro2007,citro2008,chang2009,huang2009,dalmonte2010}, superfluids, supersolids and 
stripes \cite{kollath2008,fellows2011,silva2012},
liquids of trimers and crystals of different complexes \cite{dalmonte2011,knap2012}, 
Mott insulators \cite{arguelles2007,bauer2012}, exotic quantum
criticality \cite{leche2012,tsvelik2012}, zig-zag transitions \cite{ruhman2012} 
and few-body states of several 
molecules \cite{wunsch2011,zinner2011,santos2010}
have been discussed in the literature.

\begin{figure}[ht!]
\centering
\includegraphics[scale=0.6]{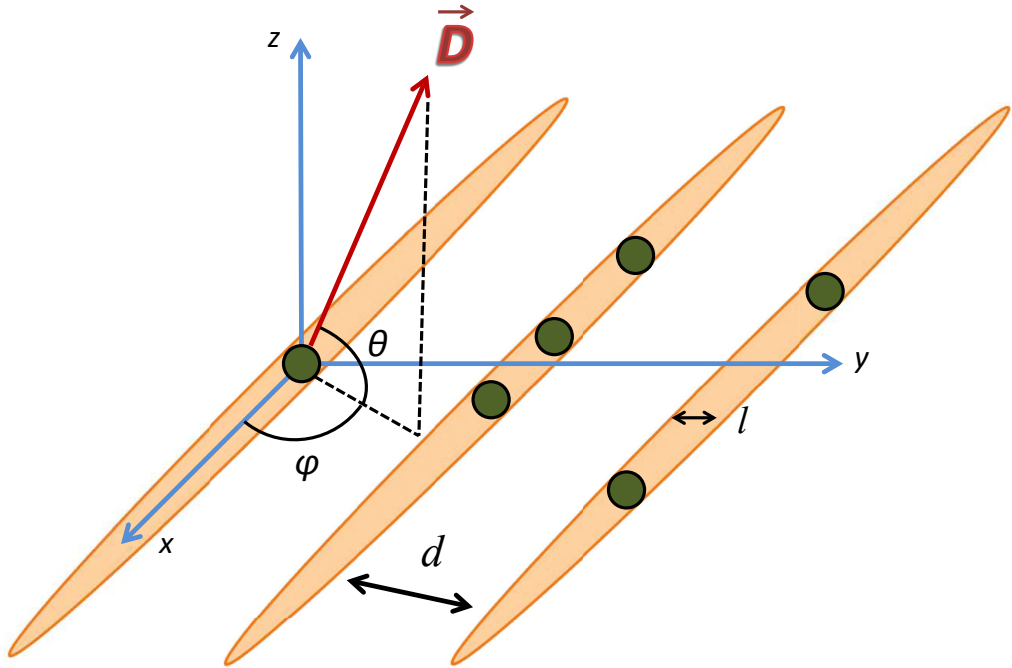}
\caption{Schematic of the setup for the case of three equidistant one-dimensional
tubes along the $x$-direction in the $xy$-plane with a distance between adjacent tubes of $d$. 
The dipolar moment, $\vec{\bm D}$, points along
the direction specified by the two angles $\phi$ and $\theta$ as defined on the figure.
Dark filled circles indicate the dipolar atoms in the tubes. The tubes have a thickness 
given by $l$. In typical experimental 
setups the tubes also have an in-tube confinement. This is indicated on the figure
by the shrinking of the tubes at both ends. This in-tube confinement will, however, 
be neglected in this work.}
\label{schematic}
\end{figure}

In the current paper we will address the few-body structures that can be expected in 
an array of several tubes. A schematic setup for three tubes is shown 
in figure~\ref{schematic}. The orientation of the dipolar molecules  in the tubes
can be controlled by an external electric or magnetic field depending on whether
magnetic atoms or heteronuclear molecules are used. This opens up a host of 
interesting phenomena since the potentials of two dipoles in a single tube and
the potential between two dipoles in different tubes change magnitude and 
sign as one changes the angles shown in figure~\ref{schematic}.

Here we are concerned with the important question of existence of $N$-body
bound states in multi-tube geometries. In order to address this we need to 
carefully consider the limits of small dipole moment and we consequently
develop a perturbative approach to weakly bound states in one dimension that
can handle the dipolar potential. At stronger coupling we consider a 
harmonic approximation. Both are compared to numerical 
results from the stochastic variational method
that allow us to test the accuracy of the analytical formalism. 
A comparison is also made to exact results for $N$-body
bosonic systems interacting through zero-range attractive forces, and 
we ask whether dipolar systems in one dimension approach
the exact results in given limits and can in turn be effectively 
described by (properly renormalized) zero-range interactions. This is 
particularly important for many-body studies since zero-range interactions
are not only convenient to work with but also often allow analytical 
results to be obtained. We also consider the two-body scattering 
dynamics of the dipoles and compare to zero-range results. Finally, 
we discuss the impact of our results for the many-body physics of 
dipolar bosons in one-dimensional geometries.

\section{Basic setup}\label{setup}
We consider the setup depicted schematically in figure~\ref{schematic}, i.e. an array
of equidistant one-dimensional tubes containing dipolar particles with 
dipole moments aligned by an external field that does not interfere with the tubular
geometry. The dipolar particles are identical bosons with mass $m$ and
dipole moment $D$. 
The potential between two dipoles with coordinates $(0,0,0)$ and $(x,nd,0)$, $V(n,x)$, 
can be written in the form
\begin{equation}
\frac{md^2}{\hbar^2}V(n,x)=U\frac{\frac{x^2}{d^2}+n^2-
3\cos(\theta)^2[\frac{x}{d}\cos(\phi)+n\sin(\phi)]^2}{(\frac{x^2}{d^2}+n^2)^{5/2}}, \label{int}
\end{equation}
where $d$ is the distance between two adjacent tubes and $n$ is an integer such that $nd$ 
is the intertube distance 
($n=1$ for nearest neighbour (adjacent) tubes, $n=2$ for next-nearest neighbours and so on).
Here we have 
defined the dimensionless dipole coupling strength, $U=\frac{mD^2}{\hbar^2d}$, where
$D$ is the dipole moment (absolute value of the vector $\vec{\bm D}$ in figure~\ref{schematic}).
For $n=0$ we get the intratube interaction in the 
limit where the tubes are strictly one-dimensional. 
In experimental setups, arrays of one-dimensional tubes are constructed by applying
optical lattices to the dipolar gas \cite{miranda2011,chotia2012}. In this case the 
tubes are not strictly one-dimensional but will have some width along the transverse
direction that is determined by the laser intensity. This can be translated into 
a gaussian wave packet in the transverse direction that will be increasingly localized
in space as the laser intensity increases. An effective interaction that takes this
into account by integrating out a gaussian wave packet can then be obtained \cite{zinner2011}, 
which will in the limit of zero gaussian width reduce to the expression in (\ref{int}).
We will assume that the lattice is very strong so that the strict one-dimensional 
expression above is valid for the interaction of particles in different tubes which 
is accurate when the transverse width, $l$, is much smaller than the intertube 
distance, $d$. Corrections to this picture have been discussed in references
\cite{wunsch2011} and \cite{zinner2011}. Below we will return to the question of finite
transverse width when we treat bound states with more than one particle per tube.
From here on we will adopt units $\frac{\hbar^2}{md^2}$ and $d$ for all energies and 
lengths. 

The potential above has the interesting property that for the case of $n\geq 1$
\begin{equation}
\frac{md^2}{\hbar^2}\int_{-\infty}^{\infty}V(x,n)dx=\frac{2U}{n^2}\left[\cos^2(\theta)\cos^2(\phi)-\cos(2\theta)\right],
\label{intpot}
\end{equation}
and as a function of the angles we thus see that we can obtain both positive and negative
values of the integrated interaction. The criterion for the existence of a two-body bound 
dimer in the limit of small $U$ is a negative integral \cite{landau1977,simon1976}. We thus
see that one can control the presence of a weakly-bound dimer by changing $\phi$ and $\theta$. Below
we will mostly address the situation where $\phi=\pi/2$ and $\theta=0$, i.e. dipoles
that are oriented perpendicular to the tubes. In this case, the integral in (\ref{intpot})
is $-\frac{2U}{n^2}$ and this means that the system can produce bound dimers between any
two particles that are located in different tubes for any $n>0$. 

However, 
if a given few-body state has more than one particle in a single tube, then the interlayer
interaction will be either attractive or repulsive depending on the sign
of $1-3\cos^2(\theta)\cos^2(\phi)$ (see (\ref{int}) with $n=0$). 
Here we will only consider the regime where
this quantity is positive, i.e. the case for which two dipoles in a single tube
repel each other in order to avoid any collapsing states within the tubes.

\section{Few-body bound states}
We now present our results for up to five dipolar bosonic
particles in multi-tube geometries. The discussion will involve analytical tools
for addressing the limits of weak and strong dipolar interactions which will be 
compared to numerics using the exact stochastic variational 
approach \cite{artem2011a,artem2011b}. This allows us to determine the
range of validity of the analytical methods that we employ. Along the way we 
will make a detailed comparison between the one-dimensional case with tubes
and the two-dimensional case of multiple 
layers \cite{miranda2011,wang2006,wang2007,jeremy2010,klawunn2010,baranov2011,artem2012,zinner2012}.
We will also compare to the results of McGuire \cite{mcguire1964} for the exact
ground state energy of an $N$-boson system with pairwise zero-range interactions
in one dimension. Finally, we will investigate few-body states with more than one
particle in a single tube for both the case of perpendicular dipoles ($\theta=0$ and $\phi=\pi/2$) 
and with tilted angles ($\theta=0$ and $\phi<\pi/2$).

\subsection {Two-body states of two dipoles in two tubes}
The first case is one dipolar particle in each of two adjacent tubes with dipole moments
oriented perpendicular to the tubes ($\phi=\pi/2$ and $\theta=0$). 
This configuration has the Schr{\"o}dinger equation
\begin{equation}
\left(-\frac{\partial^2}{\partial x^2}+U\frac{x^2-2}
{(x^2+1)^{5/2}}\right)\Phi=\epsilon \Phi,\label{shr-eq}
\end{equation}
where $x$ is the relative distance of the two dipoles along the 
tube axis, $\epsilon$ is the eigenenergy in units of $\frac{\hbar^2}{md^2}$ and
$\Phi$ is the wave function. We start with the analytically accesible limits 
of strong $\epsilon/U \sim 1$ and weak $\epsilon/U \ll 1$ 
binding.   

{\it Weak binding.} From reference \cite{simon1976} we know that a bound state 
exists for potentials that fulfil $\int (1+|x|)V(x)dx<\infty$. For this bound state we 
write the solution of (\ref{shr-eq})
in the form 
\begin{equation}
\Phi(\kappa,x)=e^{-\kappa x}-\int_x^\infty \mathrm{d}y 
\frac{\sinh(\kappa[x-y])}{\kappa}V(y)\Phi(\kappa,y),\label{wave}
\end{equation}
where $\epsilon=-\kappa^2,\kappa>0$
\cite{newton1986}. Notice that we are 'normalizing' the wave function 
at $x=\infty$ and the integrating our way to any other value of $x$. 
For $x<0$ the term $e^{-\kappa x}$ appears to diverge, but the second 
term will also contribute and the combination yields the correct answer.
We now obtain
the equation for the binding energy
\begin{equation}
\kappa=-\frac{1}{2}\int_{-\infty}^{\infty} 
\mathrm{d}y e^{\kappa y} V(y)\Phi(\kappa,y). \label{bind-en}
\end{equation}
For weak binding this can be solved iteratively for small $\kappa$, i.e.
\begin{equation}
\kappa=-\frac{1}{2}\int_{-\infty}^{\infty} V(y) \mathrm{d}y 
+\frac{1}{2}\int_{-\infty}^{\infty} \mathrm{d}y V(y) 
\int_y^{\infty}\mathrm{d}y'(y-y')V(y')+\ldots
\label{weak}
\end{equation}
Here we have used that $\Phi(\kappa,y)\approx 1+O(\kappa)$.
Upon inserting the potential in (\ref{int}) with $n=1$ and carrying out the 
integrals we find
\begin{equation}
\kappa=
U+U^2\frac{\pi}{16}+o(U^2),
\end{equation}
which implies that to leading order in $U$, $\epsilon=-\kappa^2\propto -U^2$. 
These results were initially discussed by Landau \cite{landau1977} 
and Simon \cite{simon1976} but
the procedure we present here allows simpler access to the higher-order
terms. We note that the first term in (\ref{weak}) corresponds 
to a delta-function potential 
$-g\frac{\hbar^2}{md^2}\delta(\frac{y}{d})$ with $g=2U$ (where we 
have reintroduced explicit units for clarity). Numerically we find that this potential
gives good approximation for the real potential 
with $U<0.2$. This is a valuable conclusion for studies that depend on
a delta-function representation of the dipolar interaction in one
dimension in order to describe many-body physics \cite{leche2012}.

\begin{figure}[ht!]
\centerline{\epsfig{file=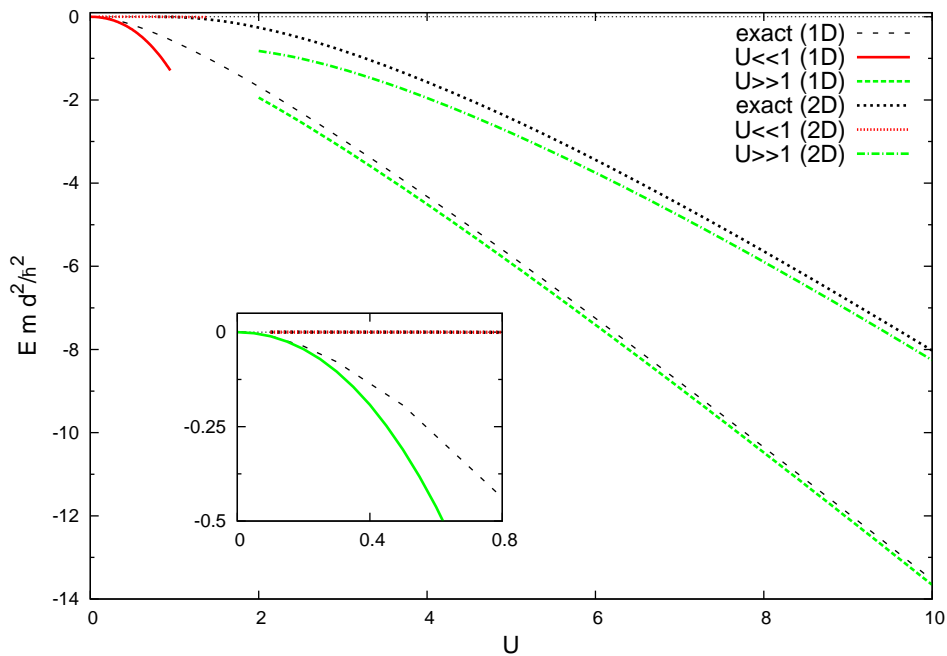,scale=1.0}}
\caption{Energy, E, in units of $\frac{md^2}{\hbar^2}$ as a function of 
dimensionless dipolar strength, $U$.
Comparison between exact numerical solution 
of the Schr{\"o}dinger equation (\ref{shr-eq}) and analytical 
predictions from table~\ref{tab} for 1D and 2D. The inset is a magnification 
of the small $U$ regime. Note that the huge difference at small $U$ between 
1D and 2D due to the exponential decrease of the binding energy at small 
coupling strength in 2D.}
\label{figure1}
\end{figure}

A clear difference to the 2D case is seen at this point. The 
weak-binding result in 1D is $\epsilon\propto -U^2$, whereas in 2D 
it is $\epsilon\propto -\exp(-8/U^2)$ \cite{artem2011a,jeremy2010,klawunn2010,baranov2011}.
This is consistent with our expectation that weakly-bound 1D dipolar few-body bound states
are generally more stable than similar systems in 2D, and any sort
of external perturbations would have a less severe effect in the 1D
setup. 

We can also consider the case where $U<0$ corresponding to one of the 
dipoles in the two tubes pointing in opposite directions which should
be possible to achieve using AC fields \cite{neyenhuis2012}. In this
case the 2D setup with two adjacent planes will {\it always} have a 
bound dimer \cite{artem2011b}, whereas in 1D we find numerically 
that it takes a finite strength $U<-4.98$ to bind the two-body
system.  We can understand the latter if we consider $\kappa=0$
in (\ref{bind-en}). This
transforms the condition for the appearance of the 
bound state. For small $U$ we have 
$U+U^2\frac{\pi}{16}=0$, which has two solutions;
one corresponds to bound states with $U>0$
and another for bound states with 
$U<-\frac{16}{\pi}\sim -5.09$. The latter is
within 2 percent of the numerical value. However, notice that
the $U<0$ case opens up the possibility of binding and unbinding the 
dimer by tuning $U$ as originally envisioned in reference \cite{wang2006}.

{\it Strong binding 1D}
For this case we may assume that our wave function is 
strongly localised near the origin, i.e.
$\langle x^2 \rangle / \epsilon \rightarrow 0$, such that the probability
to be in the classically forbidden region is small.
Under these conditions we can use standard techniques to 
obtain the binding
energy and the wave function. 
First we decompose the potential near the origin 
\begin{equation}
U\frac{x^2-2}{(x^2+1)^{5/2}}=-2U+6Ux^2-45Ux^4/4+U\sum_{n=3}^\infty \alpha_n x^{2n}.
\end{equation}
Using the assumption of strong localization 
we see that up to terms $O(\frac{1}{\sqrt U})$
we can solve the harmonic oscillator problem for the 
first two terms and include the third term via perturbation 
theory. This yields
\begin{equation}
\epsilon = -2U+\sqrt{6U}-45/32 +O(\frac{1}{\sqrt{U}}).
\end{equation}
A comparison of the numerical results for all values of $U$ to the weak and 
strong binding expansions is shown in figure~\ref{figure1}. Both the limit
of weak and strong binding are well reproduced by the different approximation 
schemes. We see differences between numerical and analytical results only around $U=1$.
For completeness
we also summarize all the analytical results and their regimes of validity
in table~\ref{tab}

\begin{table}[ht!]
\centering
\begin{tabular}{ | l | c | }
\hline
   & 1D  \\
\hline
  $\frac{md^2}{\hbar^2}\int V \mathrm{d}\mathbf{r}$ & -2U  \\
\hline
  small $U>0$ & $\epsilon=-U^2-\frac{\pi}{8} U^3+o(U^3), (U<0.2)$   \\
\hline
  large $U>0$ & $\epsilon=-2U+\sqrt{6U}-45/32+O(\frac{1}{\sqrt U}), (U>4)$ \\
\hline
  & 2D\\
\hline
  $\frac{md^2}{\hbar^2}\int V \mathrm{d}\mathbf{r}$ &  0 \\
\hline
  small $|U|>0$ & 
$\epsilon=-4 \exp({-2\gamma-\frac{8}{U^2}+
\frac{128}{15 U}-\frac{2521}{450}+o(U)}) , (U<0.8)$ \\
\hline
  large $U>0$  
& $\epsilon=-2U +\sqrt{24U}-15/4 + O(\frac{1}{\sqrt U}), (U>7)$ \\
\hline
\end{tabular}
\caption{Comparison between 1D and 2D binding energies of two dipoles in 
two different layers or tubes with perpendicularly oriented dipole moments. The 
regimes of validity of the different analytical formulae are indicated in the parenthesis.}
\label{tab}
\end{table}

\subsection{$N$-body Chains}
We now proceed to discuss the case where we have $N$-body chains that 
consist of one dipolar particle in each of $N$ adjacent layers or tubes.
In the 2D case, these sorts of structures have generated a lot of 
recent interest since one expects highly non-trivial few- and many-body dynamics
\cite{wang2006,jeremy2010,potter2010,pikovski2010,armstrong2011,barbara2011,armstrong2012,zinner2012b}.
There is a similar interest in the one-dimensional multi-tube configurations 
\cite{fellows2011,arguelles2007,leche2012,tsvelik2012}. The Hamiltonian for 
the $N$-body 1D system is 
\begin{equation}
H=\sum_{i=1}^{N} \frac{p_{i}^{2}}{2m}+\sum_{i>j}V(x_{ij},n_{ij}),\label{hamil}
\end{equation}
where $x_{ij}$ is the relative distance of the $i$th and $j$th particles 
along the tube direction and $n_{ij}$ is the integer that signifies how far
apart the tubes are that hold the $i$th and $j$th particles ($n=1$ for 
adjacent tubes and so on). For the case where the pair-wise potentials
of equal strength and zero-range, 
this Hamiltonian was studied in the classical papers of 
Lieb and Liniger \cite{lieb1963a,lieb1963b}. McGuire has shown that 
for this delta-function case, the bound state energy, $E_N$, of the $N$-body problem
can be written very elegantly as $E_N/E_2=N(N^2-1)/6$, where $E_2$ is the 
two-body bound state energy of the same delta-function potential \cite{mcguire1964}. 

Here we consider a system of $N$ dipoles placed in $N$ different 
tubes with the potential in (\ref{hamil}) given by (\ref{int}). This 
is intrinsically a case with long-range interactions which has also 
generated some classical studies. In the particular case where the 
interaction behaves as $r^{-2}$ this is the 
Sutherland-Calogero model \cite{suther1971,calogero1971}.
In the present case with dipoles we have a long-range tail that behaves as $r^{-3}$.
We would like to build a connection to the zero-range studies for the 
dipolar systems in 1D. For general polarization angles, however, the 
interactions are anisotropic and more difficult to compare to the 
isotropic zero-range models. In this section we therefore study the 
case of perpendicular polarization ($\phi=\pi/2$ and $\theta=0$) 
where the anisotropy is absent.

\begin{figure}[ht!]
\centerline{\epsfig{file=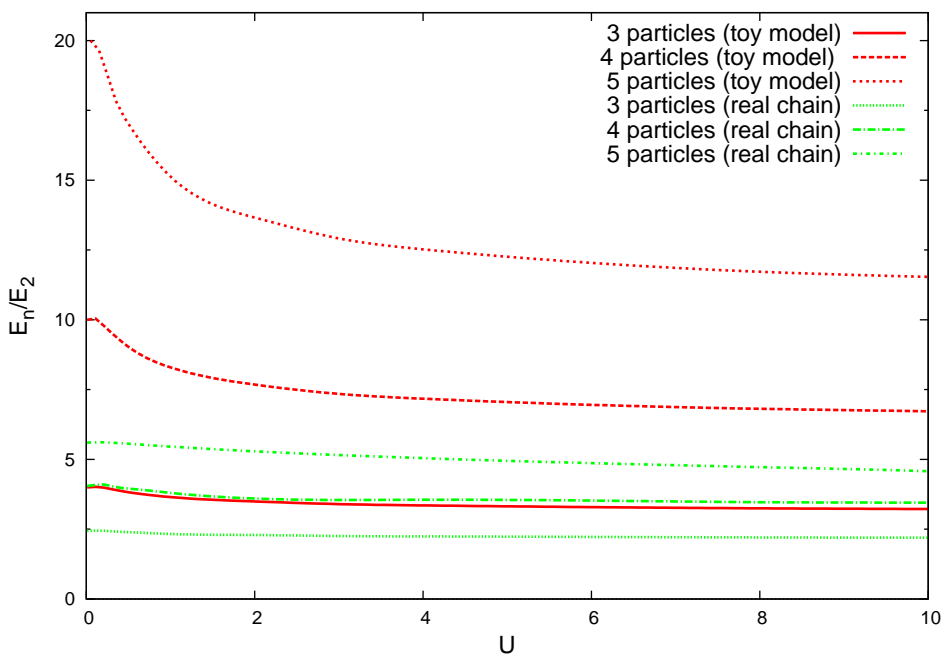,scale=1.0}}
\caption{Numerically determined $N$-body binding energies relative to the 
two-body energy, $E_2$, as function of coupling strength, $U$, 
for $N$-body chains using both a toy model where all 
interactions have $n=1$ in (\ref{int}) and the realistic case where $n$ 
is different for different pairs of particles in the chain.}
\label{figure2}
\end{figure}

As we have demonstrated above, for $U<0.2$
we get universal behaviour in a sense that the potential 
$\frac{md^2}{\hbar^2}V=-2U\delta(\frac{x}{d})$ accurately describes our system. 
Now we want to establish how well this zero-range 
approximation works on the few-body level with $N>2$.
To do so we first consider a
toy model where all $N$ particles interact with the same 
interaction corresponding to $n=1$ in (\ref{int}). This 
is precisely the case where it makes sense to compare to the 
analytical result
$E_N/E_2=N(N^2-1)/6$. In this way we 
can get a feeling for when the zero-range approximation works.
For realistic chains $n$ is not always equal to one. This 
means that these are easier to describe using a 
zero-range approximation since particle
pairs that are located several tubes apart in a chain have a potential 
that effectively correspond to a smaller value of $U$ (as compared to 
the $n=1$ case). 

The numerically calculated 
chain binding energies relative to the two-body energy, $E_N/E_2$, 
for both the toy model and the real dipolar
chains in 1D are presented in figure~\ref{figure2}. 
We observe the delta-function behaviour 
in all cases for $U<0.15$, and the strong coupling behavior
is slowly approached with increasing interaction strength. In current
experiments with $^{40}$K-$^{87}$Rb molecules \cite{miranda2011,chotia2012}, 
we have $U\lesssim 0.1$ so this is weak-coupling in the current context.
In the strong coupling limit, the leading term for the toy model  
is $E_N/E_2=\frac{N(N-1)}{2}$ which is simply the number of 
pairs that all contribute an energy $E_2$. For the realistic chain, 
the expression is instead
\begin{equation}\label{strong}
E_N/E_2=\sum_{k=1}^N \sum_{i=1}^{N-k}\frac{1}{i^3}=NH_{N-1}^{(3)}-H_{N-1}^{(2)}, 
\end{equation}
where $H_{m}^{(k)}=\sum_{i=1}^{m}i^{-k}$ is the harmonic number of order $k$.
We will elaborate more on the strong-coupling limit below.

We see a large difference between the toy model and the real chain
in figure~\ref{figure2} in the weak-coupling limit $U\ll 1$. This is 
an important conclusion of our work that has implications for 
many-body studies on dipolar systems that work with delta-function 
approximations for the dipolar interactions. The toy model 
is relevant for an equidistant triangular tube configuration as
studied in reference~\cite{leche2012}. Our results for this 
$N=3$ case show a very flat profile for all values of $U$ and 
from the point of view of energetics it is thus a reasonable 
approximation to use the delta-function $\frac{md^2}{\hbar^2}V(x)=-2U\delta(\frac{x}{d})$.
The $N=4$ and $5$ would be applicable to a tube configuration with
four and five nearest-neighbours respectively (the latter would
of course not be possible with standard crystal lattices). 
However, in these
cases the variation of the energies with $U$ is much more drastic, 
indicating that a zero-range description is only good for very small 
$U$. 

For the real chain systems, the curves are rather flat for all $N$.
This is connected to the fact that the terms with $n>1$ in (\ref{hamil})
will give a contribution that is suppressed. Due to the flatness of 
the curves in figure~\ref{figure2}, we can get a good approximation for
the $N$-body binding energy of the realistic chains by using an 
effective model for $N$-body systems 
with $\frac{md^2}{\hbar^2}V(x)=-2U_{eff}(N)\delta(\frac{x}{d})$ where
\begin{equation}
U_{eff}(N)=\frac{6}{N(N^2-1)}\left[NH_{N-1}^{(3)}-H_{N-1}^{(2)}\right]U.
\label{Ueff}
\end{equation}
The blessing of working with a zero-range interaction comes at the price
of having an $N$-dependent coupling. This ensures that the $N$-body 
bound states present in this geometry are properly described. We will 
use this effective dipolar strength to study the corresponding many-body 
system below. 
We stress that the delta-function approximation 
with $U_{eff}(N)$ coupling is only accurate for weak interaction, i.e.
small $U$. In the case of $N=5$ from $U\to 0$ and up to $U=10$ we find 
a difference of about $20\%$ on the value while 
for $N=3$ and $N=4$ the difference is slightly less. So the flatness and thus the 
reliability of the delta-function approach is on the level of $20\%$
for $U\leq 10$ and $N\leq 5$. It is very important to notice that we
reproduce the strong-coupling limit, i.e. large $U$, with this choice
of $U_{eff}$. This means that if we use the potential in (\ref{Ueff})
as an effective potential between $N$ bosonic particles in one 
dimension, we obtain the correct strong-coupling binding energy (\ref{strong}).
This all builds on the flatness of the energy in figure~\ref{figure2}
and is only as good as the uncertainty quoted above.
This is important to notice as below we will look at low-energy 
many-body physics using this effective interaction.

\begin{figure}[ht!]
\centerline{\epsfig{file=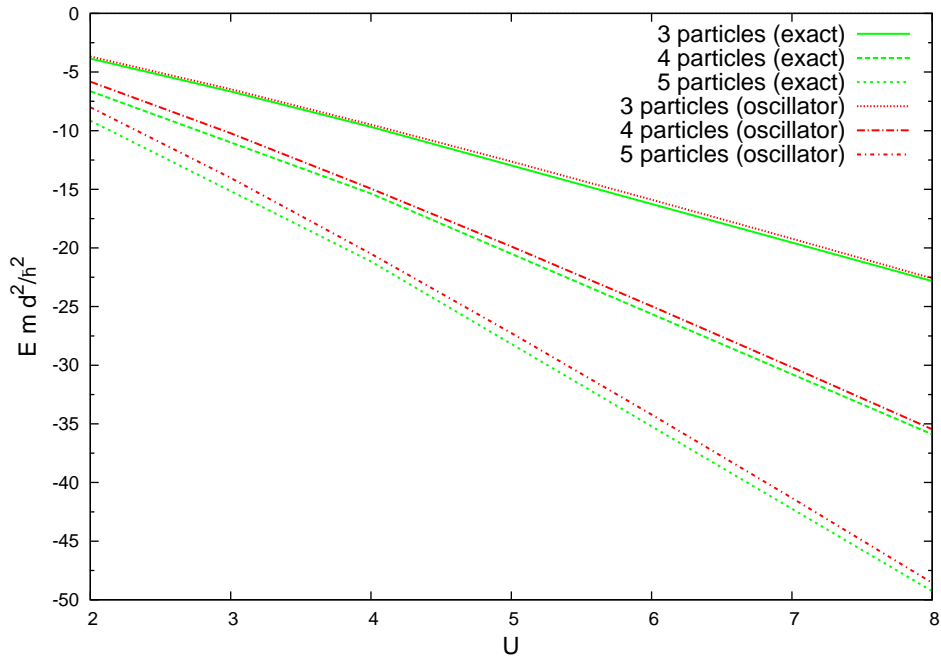,scale=1.0}}
\caption{Energy, E, in units of $\frac{md^2}{\hbar^2}$ as a function of 
dimensionless dipolar strength, $U$. 
Comparison between exact numerical calculations obtained from the 
stochastic variational approach and the oscillator approximation described in the
text for the case of perpendicular dipoles, i.e. 
$\phi=\pi/2$ and $\theta=0$ for the case of $N=3$, $4$ and $5$.}
\label{figure3}
\end{figure}

 {\it Chains in the oscillator approximation}
As we have seen for small $U$, the 
zero-range approximation gives reliable results. However, 
we would also like to discuss the strong binding limit where
$U\gg 1$ and provide a simple analytical procedure for 
obtaining the energies and wave functions. This can 
be done using exact harmonic models \cite{jeremy2011,jeremy2012} 
where the full two-body
interaction (the dipolar interaction in this case) is  
represented by a harmonic oscillator with parameters that 
are carefully chosen to reproduce energetics and structure
at the two-body level \cite{armstrong2012}. We expect this to be a
good approximation for $U\gg 1$ since the potential in (\ref{int})
has a deep pocket in this case. However, it turns out that this
approximation can be very accurate even for moderate $U$ in
the case of 2D multi-layered systems with chains \cite{artem2011c}.

In figure~\ref{figure3} we show a comparison of the oscillator
approximation to exact numerical calculations for the case
of perpendicularly oriented dipoles ($\phi=\pi/2$ and $\theta=0$).
Note that the horizontal axis begins at $U=2$ (for $U<2$ the 
oscillator does not reproduce the exact results well).
We get a very good agreement for $N=3$, while for higher particle
numbers there is a clearly visible deviation. This stems from 
issues with the two-body interactions coming from the outermost tubes 
where the oscillator approximation starts to fail since the contributions
are not well-described by an oscillator (the effective $U$ for these 
terms is small for any $U$ value considered here). However, the 
overall agreement is still within a few percent for $U>5$. One way to improve
the agreement for higher particle numbers could be to introduce
an effective three-body force (more generally an $N$-body force)
in addition to the two-body potential terms. Such an approach is
very common in nuclear physics and in effective field theory 
studies of universal three-body physics \cite{hammer2012,frederico2012}.

\begin{figure}[ht!]
\centerline{\epsfig{file=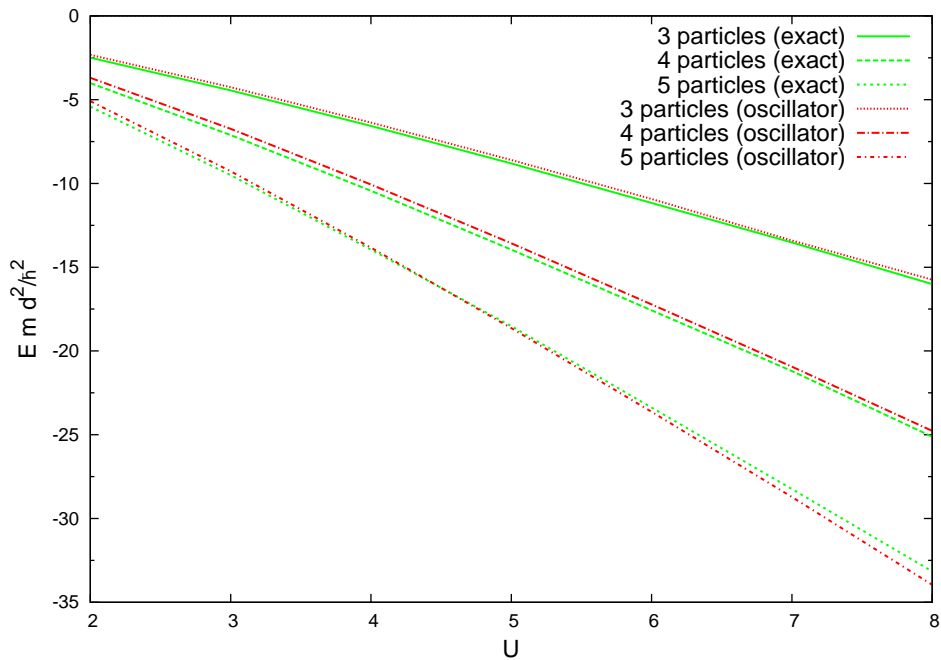,scale=1.0}}
\caption{Energy, E, in units of $\frac{md^2}{\hbar^2}$ as a function of 
dimensionless dipolar strength, $U$.
Same as in figure~\ref{figure3} but for the case
where $\phi=\phi_m$ and $\theta=0$. This corresponds to vanishing 
intratube interaction, i.e. two dipolar particles in the same tube do 
not interact.}
\label{figure4}
\end{figure}

We now change the angle $\phi$ to investigate the effects of 
anisotropy. Figure~\ref{figure4} shows the case with 
$\phi=\phi_m$ and $\theta=0$, where $\phi_m$ satisfies
$\cos^2\phi_m=\frac{1}{3}$. Recalling the discussion after 
(\ref{int}) this implies that the potential between two
dipoles in the same tube vanishes.
First we observe that the overall energy scale becomes smaller,
because the depth of the potential and net volume of the 
potential is smaller than for 
the perpendicular polarization case in figure~\ref{figure3}. 
For $N=3$ and $N=4$ we again see very good agreement between
the oscillator approximation and the numerics. The largest
chain with $N=5$ shows what appears to be an even better
agreement than for the $\phi=\pi/2$ and $\theta=0$ case.
What is also striking is that the oscillator results and 
the numerics cross around $U\sim 5$. This happens since
our method ensures that the oscillator reproduces 
the slope of the numerical results in the limit $U\gg 1$
(as discussed for the 2D case in reference~\cite{artem2011c}).
The optimization of the parameters for the two-body 
oscillator potential for $2\leq U\lesssim 5$, however, 
produces a slightly different slope and a crossing 
thus takes place.

\subsection{Non-chain bound complexes}
We now consider the case where there are two particles
in one tube. The simplest configuration of this sort is 
two tubes with three particles in total, i.e. one particle 
in one tube and two particles in the adjacent tube. 
At this point we need to take the 
width of the tube that we discussed in section~\ref{setup}
explicitly into account in order to work with an intratube
interaction that best describes the experimental situation.
This can be done by folding the potential with a gaussian 
wave packet of width $l$ that represents the wave function of
the system in the transverse direction (along the $y$-axis in figure~\ref{schematic})
This yields the intratube interaction \cite{zinner2011}
\begin{equation}
V_{rep}(x)=U_{rep}\lambda^3(1-3\cos^2 \phi \cos^2 \theta)f(\lambda x) ,
\label{vrep}
\end{equation} 
where  
\begin{equation}
f(z)=\frac{-2|z|+\sqrt{2\pi}(1+z^2)\exp(z^2/2) 
\mathrm{erfc}(|z|/\sqrt{2})}{4},\label{fold}
\end{equation} 
and $\lambda=d/l$ and $U_{rep}$ is the 
strength of the intratube interaction. 
This means that the intratube potential has been regularized 
by the finite width of the transverse confinement and does
not have the strict 1D form $x^{-3}$ {\it expect} in the 
limit of $\lambda\to\infty$. 
In a system where the 
dipole moment is the same for all particles in all tubes we have 
$U_{rep}=U$, where $U$ is the strength of the intertube interaction. 
The dipole orientation and the induced dipole 
moment is typically controlled by applying an external electric or 
magnetic field that is constant across the whole system. However, 
if one applies also a field gradient then it should be possible to 
obtain a system where the induced dipoles moment varies from tube
to tube (or layer to layer in the 2D multilayer case). In this 
case we can have a situation where $U_{rep}\neq U$ and 
we thus vary both quantities independently in the current study.
As discussed previously, we 
will consider only the repulsive case where 
$1-3\cos^2\phi\cos^2\theta>0$ in order to avoid possible intratube
collapse of the system. Unless explicitly states, we use $\lambda=5$ 
everywhere (see reference \cite{zinner2011} for a discussion of the 
effects of changing $\lambda$).

We start by considering the case where all particles are 
perpendicularly polarized, i.e. $\phi=\pi/2$ and $\theta=0$.
Observe that if $U_{rep}=0$ we have a trimer 
energy that is below the dimer energy, 
and if $U_{rep}\to \infty$ we can not bind three particles. 
We now want to find the value $U_{rep}^{cr}$ at which the 
three-body system is bound for a given value of $U$, so 
we keep $U$ fixed and vary $U_{rep}$.
In reference~\cite{artem2012}
it has been shown that $U_{rep}^{cr}/U$
is an increasing function of $U$ by an argument that does
not depend on the dimensionality and thus applies equally 
well to layers in 2D and tubes in 1D.
In turn we only need to consider
$U\rightarrow\infty$ to obtain an 
upper bound for $U_{rep}^{cr}/U$.
This upper bound is a decreasing function of $\lambda$ 
and for
$\lambda=5$ we get $U_{rep}^{cr}/U<0.41$. 
In order to find the 
lower bound for $U_{rep}^{cr}/U$ we will use the variational 
principle with a normalised three-body wave function of the form
$\Psi(x,y)=\Phi(x)\varphi(y)$, where $\Phi(x)$ is the 
normalised dimer wave function and we have used Jacobi 
coordinates $\sqrt{2}x=x_1-x_2$ and $\sqrt{3/2}y=x_3-(x_1+x_2)/2$. 
From this we obtain the upper bound for the three-body energy
\begin{eqnarray}
&E_3^{var}=\langle \Psi|H_3|\Psi\rangle=E_2+&\nonumber\\
&\langle\varphi|\bigg(T_y+\int \Phi^2(x)
\bigg[V_{rep}(|\sqrt{\frac{1}{2}}x+\sqrt {\frac{3}{2}}y|)+V(|\sqrt{\frac{1}{2}}x-
\sqrt {\frac{3}{2}}y|)\bigg]\mathrm{d}x\bigg)|\varphi\rangle,&\label{e3var}
\end{eqnarray}
where the kinetic term for the $y$ coordinate 
is $T_y=-\frac{\hbar^2}{2m}\frac{\partial^2}{\partial y^2}$.
Here $V_{rep}$ is given in (\ref{vrep}) and $V$ is given
by (\ref{int}) with $n=1$.
Equation (\ref{bind-en}) in principle allows us to estimate 
where $E_3^{var}<E_2$ holds, which gives us the lower bound 
for $U_{rep}^{cr}$. However, in general this equation
is not easy to analyse so we
obtain the lower bound using the simpler 
condition for the effective potential acting on 
$\varphi$ in (\ref{e3var}) which is that it must have
negative net volume. The limit of zero net volume 
requires
\begin{equation}
\int \int \Phi^2(x)\bigg[V_{rep}(|\sqrt{\frac{1}{2}}x+
\sqrt {\frac{3}{2}}y|)+V(|\sqrt{\frac{1}{2}}x-
\sqrt {\frac{3}{2}}y|)\bigg]\mathrm{d}x \mathrm{d}y=0,
\end{equation}
which can be simplified to 
\begin{equation}
 \int \bigg[V_{rep}(y)+V(y)\bigg]\mathrm{d}y=-2U+U_{rep}
\lambda^2\int_{-\infty}^{\infty}f(y)\mathrm{d}y=-2U+\lambda^2U_{rep}=0.
\end{equation}
This lower bound is a decreasing function of $\lambda$, 
and for $\lambda=5$ we get $U_{rep}^{cr}/U>2/25$. Collecting 
our results, we have for $\lambda=5$ that
$0.41>U_{rep}^{cr}/U>2/25$. This means that 
if we could manipulate the values of the 
intratube and intertube interactions independently,
we could create or destroy the three-body bound
state in two tubes with perpendicular polarization. 
This effect is completely due to the finite range of 
the potential, because for zero-range potentials
any value $U_{rep}/U$ will produce a bound state 
(the system becomes equivalent to the fermionic states
studied by McQuire in reference~\cite{mcguire1966} for which 
a bound state always exists).

\begin{figure}[thb]
\centerline{\epsfig{file=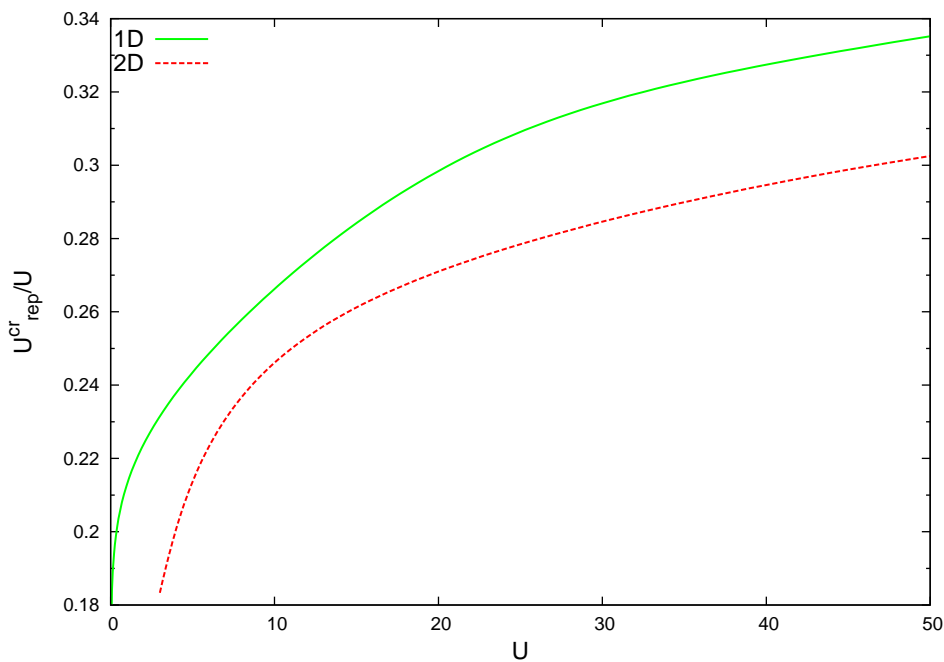,scale=1.0}}
\caption{Ratio $U^{cr}_{rep}/U$ as a function of 
U at which a three-body state with two dipoles in 
one tube (plane) and one in the adjacent tube (plane) becomes
possible in 1D (2D). The three-body state exists below the 
lines in the plot. The lines drops sharply to zero for small 
$U$. Here we use $\lambda=5$.}
\label{figure5}
\end{figure} 

In figure~\ref{figure5} we show the critical value, $U_{rep}^{cr}/U$, 
as a function of $U$ for both the case of 1D tubes and for comparison 
we also plot the results of the same study for a bilayer in 2D \cite{artem2012}.
Both of these calculations have been done with $\lambda=5$. We see a clear
difference between the critical values for three-body bound state formation 
in the two cases with the 1D situation allowing for larger repulsive 
interactions. This is connected to the price one pays for localizing particles
in a bound states complex. In 2D we need to localize in two independent directions
which is more expensive than 1D. As expected, 1D tubes are more favourable for 
formation of these larger complexes.

Having discussed how a tuning of the
intratube repulsion can bind a three-body state, we 
now consider an alternative route to bind this state which 
is by tilting the angle, $\phi$. We assume 
that $U_{rep}=U$, and are looking for values of $\phi$ and $U$
for which a bound state exists.
For a given angle $\phi$ we introduce the notation $U^{cr}(\phi)$ which
is the value at which the two-body dimer (one particle in each 
of two adjacent tubes) energy is equal to the three-body 
energy which defines the threshold.
We start with three 
particles as before but 
tilted by angles $\theta=0, \phi \in [\phi_m ,\pi/2]$.
We already proved that 
for $\phi=\pi/2$ we do not have a bound state if 
$U_{rep}=U$. However, we know that for $\phi=\phi_m$
we always have a three-body bound state, because 
the intralayer interaction vanishes. This implies 
that there is a range of angles where
the three-body state exists.

To approach this problem analytically, we can calculate
the net 
volume as above. This yields
 \begin{equation}
 \int \bigg[V_{rep}(y)+V(y)\bigg]\mathrm{d}y=2U(\cos^2\phi-1)
+U(1-3\cos^2\phi)\lambda^2.
\end{equation}
Putting it to zero we obtain the smallest angle below 
which the three-body bound state exist for any value of $U$.
The condition for a three-body bound state is 
$\phi\leq\cos^{-1}(\sqrt{\frac{\lambda^2-2}{3\lambda^2-2}})$ which
in the case of $\lambda=5$ yields $\phi\lesssim 0.975$ or $55.9^o$. 
In the strict 1D limit where
$\lambda\to\infty$, we find the result $\phi\leq\phi_m$.
The function $\frac{\lambda^2-2}{3\lambda^2-2}$
is decreasing with $\lambda$ which is a result of the 
behaviour of the function in (\ref{fold}) which has a 
finite value for zero argument for any finite $\lambda$. 
In strict 1D we expect
that the repulsion will not allow this three-body bound state 
due to the 
cubic singularity $x^{-3}$ of the repulsive
potential. For the more realistic case with a finite
$\lambda$ we can have bound states for angle in the regime
$\phi>\phi_m$ as well.
An additional bound can be found by using classical arguments,
i.e. neglecting kinetic energy in the Schr{\"o}dinger equation 
and studying just the potential landscape \cite{artem2012}. 
This yields a bound of $\phi\gtrsim 1.15$ or $66^o$ so that 
for any $U$ we 
do not have three-body bound states in the case where $U_{rep}=U$.
In summary, we have found that for
$\phi_m<\phi\lesssim 1.15$ there exist a value of $U^{cr}(\phi)\in [0,\infty)$,
such that for $U>U^{cr}(\phi)$ the bi-tube trimer is bound. 

\begin{figure}[thb]
\centerline{\epsfig{file=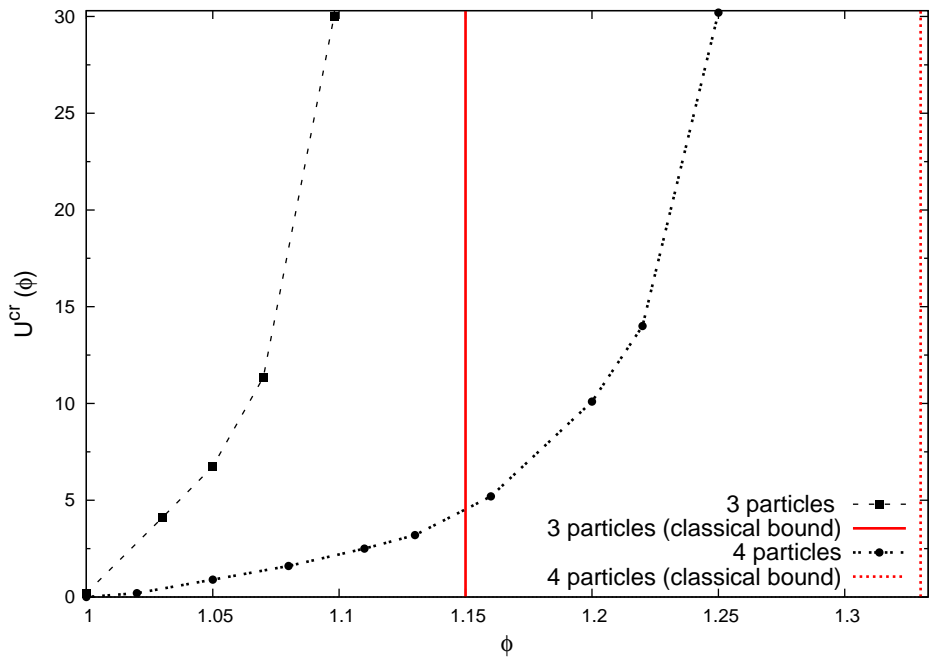,scale=1.0}}
\caption{Critical value, $U^{cr}(\phi)$, where the trimer in two
tubes (one and two particles) and tetramer in three tubes (one, two and 
one particles) dissociates as function of the tilting angle $\phi$ for
$\theta=0$. The dissociation thresholds where calculate at the 
dots and squares with connecting lines to guide the eye. The vertical
lines mark the angles beyond which a classical calculation (neglecting
kinetic energy for all particles) predicts no bound states can occur
because the system is overall repulsive.}
\label{figure6}
\end{figure} 

In figure~\ref{figure6} we plot $U^{cr}$ as a function of $\phi$ for 
three particles in two tubes in the case where $U_{rep}=U$, i.e. 
the case of current experimental interest where the dipole strength
is
controlled by one external field that is constant over the 
entire samples.
The points indicate the values at
which the dissociation points have been obtained numerically 
(connecting lines are a guide to the eye). The three-body 
bound states are present on the left-hand side in the diagram
since here the intratube repulsion is not strong enough to 
break the bound state apart.
The results obtained 
here are consistent with results reference~\cite{zinner2011} 
for $U<15$ where a comparison can be made.
For $U>15$ we find
a very sharp increase in $U^{cr}(\phi)$ that takes place
at $\phi\sim 1.1$, a few degrees before a classical 
calculation renders the system unbound at $\phi=1.15$. 

In order to address the question of additional tubes and 
their effect on bound states with more than one particle
per tube, we have also consider a four-body state in three
tubes (as in figure~\ref{schematic}) with two dipoles in 
the middle tube and one in each of the two outer tubes.
The results are also presented in figure~\ref{figure6}. 
We again see a sharp increase for large $\phi$ before
we get to the classically unbound value of $\phi=1.34$
or $76.5^o$. However, there 
is a significant increase of region in which 
bound state with two particles in a single tube are 
present due to the extra attraction from adding a 
particle in an adjacent tube on the other side
of the complex as compared to the three-body system with 
two and one in two tubes. Other four-body states
can occur in two tubes as discussed in reference~\cite{zinner2011},
but the parameter regime is severely suppressed due to the 
extra repulsion in the three-plus-one and two-plus-two 
four-body states in two tubes. We therefore expect 
that the case of three tubes could have interesting 
'quasi'-Wigner crystal states for large $U$ beyond
the clustered states studied in reference~\cite{knap2012} 
where the clusters are not of the same size in each 
tube (here with cluster size one in the outer tubes
and two in the inner tube). We can continue this 
kind of game for five tubes in a one-two-one-two-one
configuration and so on. 

\section{Scattering}
While we are mainly concerned with negative energy bound state structures in this 
paper, we now present a short overview of some scattering results that can be 
deduced directly based on the formalism we introduced above. One basically only
needs to make an analytical continuation of the bound state discussions above
in order to arrive at results for scattering of dipolar bosons in 1D tubes. 
Since current experiments with cold molecules aim for the low temperature 
regime where quantum degeneracy can be reached, we will focus exclusively 
on the low-energy scattering regime here.

The formal solution of the scattering problem, $\Psi$, satisfies the equation 
\begin{equation}
\Psi(k,x)=e^{ikx}-\int_x^\infty \mathrm{d}y 
\frac{\sin(k[x-y])}{k}V(y)\Psi(k,y),
\end{equation}
where the boundary condition is an out-going flux to the right with coefficient
one of the form $e^{ikx}$ \cite{newton1986}. Here we define 
the reflection, $R(k)$, and transmission, $T(k)$, coefficients
through $\Psi(k,x\rightarrow-\infty)=T(k)^{-1}e^{ikx}+\frac{R(k)}{T(k)}e^{-ikx}$ to 
be consistent with standard 1D potential scattering. By 
taking the limit $x\to-\infty$ we obtain
\begin{equation}
T(k)=\frac{2ik}{2ik-\int_{-\infty}^{\infty} \mathrm{d}x V(x)e^{-ikx}\Psi(k,x)}.
\label{trans-ampl}
\end{equation}
For $k=i\kappa, \kappa>0$, poles of $T(k)$
define bound states with the binding energy 
defined through the equation
\begin{equation}
\sqrt{(-\epsilon)}=\kappa=-\frac{1}{2}\int_{-\infty}^{\infty} 
\mathrm{d}y e^{\kappa y} V(y)\Psi(\kappa,y),\label{bind-en2}
\end{equation}
which is consistent with (\ref{bind-en}) and allows us to use
the same techniques as before to obtain iterative improvements
of the solution.

Consider now the case of dipoles oriented perpendicular to the 
layers, i.e. $\phi=\pi/2$ and $\theta=0$. In the limit of 
weak binding and low-energy scattering, $U\ll 1$ and $kd\ll U$, we find for
the transmission coefficient of two particles in 
the two different tubes
\begin{equation}
T=i\frac{k}{\kappa_b},
\end{equation}
and for the case of two particles in the same tube we 
find
\begin{equation}\label{trep}
T=-i\frac{kd}{U \lambda^2}.
\end{equation}
These results are deceptively simple, yet they contain 
all the information about the dipolar physics in one-dimensional tubes 
at low energy in the small $U$ limit. 
As expected they vanish with $k$ as $k\to 0$ which is a 
manifestation of the Wigner threshold law. Also, as $\kappa_b\propto U$
in the weak-coupling limit (see table~\ref{tab}), we see
that both expressions become large at very small $U$. This 
makes sense since the potential becomes very weak (and likewise
the derivative of the potential). We also see that in the 
strict 1D limit, $\lambda\to\infty$, the transmission is 
completely suppressed and total reflect is expected. This 
is the emergence of the impenetrable boson regime called the
Tonks-Girardeau limit \cite{tonks1936,girardeau1960}. This 
limit can be studied with dipolar bosons in a 1D setup with 
an (in-tube) harmonic trap as discussed in reference~\cite{deu2010}.

Since we would like to obtain an effective model for dipolar 
systems of $N$-chains using zero-range interactions, we need to 
relate the transmissison coeffecients above to its zero-range 
equivalent. If we write the potential as $V(x)=g\delta(x)$, we
find 
\begin{equation}
T(k)=\frac{2ik}{2ik-\frac{mg}{\hbar^2}}\to -\frac{2ik\hbar^2}{mg},
\end{equation}
for low-energy scattering. We see that we recover the $g=-2U$ result
for the case of two dipoles in different tubes 
(using our units of energy and length given in section~\ref{setup})
and we find $g=U\lambda^2$ for dipoles in the same tube.

\section{Many-Body Physics}
We now want to address the case of a many-body system of $N$ adjacent tubes 
with $M$ particles in each tube. We will focus on the case of perpendicular
orientation of the dipoles with respect to the tubes ($\theta=0$ and $\phi=\pi/2$)
and comment on the case of general angles in the end. In the perpendicular case, 
we expect no bound complexes beyond the chain structures that are bound 
states of $N$ bosons with one in each of $N$ adjacent tubes. Recent
Monte Carlo studies of the many-body system indicate that the 
longest chains possible, the $N$-chain for $N$ tubes, are the relevant 
degrees of freedom in these systems \cite{tsvelik2012,barbara2011} and 
we therefore focus our attention on these chains.

The results that we have presented thus far show that one can make an 
effective model of chains using zero-range interactions with suitably 
adjusted coupling strength as in (\ref{Ueff}) which reproduces the chain
energy in the strong-coupling limit.
This is extremely useful 
for many-body physics since a number of 1D $N$-body models with 
zero-range interactions are exactly solvable as discussed above. For 
the case of bosons with attractive zero-range interactions one in 
fact finds a gas-like state similar to the Tonks-Girardeau gas of 
repulsively interacting bosons, denoted the super-Tonks-Girardeau 
gas \cite{astra2005,batchelor2005,calabrese2007}. For both attractive and 
repulsive zero-range interactions these systems are also possible to 
describe using bosonic Luttinger liquids (see reference~\cite{cazalilla2011}
for a recent review of 1D bosonic systems). These studies have been boosted 
by recent experimental success in realizing these 1D quantum gas \cite{haller2009,haller2010}.
For dipolar bosons this 
Luttinger liquid behavior was studied by several authors 
\cite{citro2007,citro2008,dalmonte2010,kollath2008}.

Here we want to address the many-body behaviour by using the few-body 
physics information that we have derived above. For $N$ adjacent tubes
with $M$ particles in each, we can build a system of $M$ chains, and 
then we may study the effective inter-chain interactions. Using a 
zero-range interaction model allows us to map the problem into the 
realm of exactly solvable 1D systems and/or Luttinger liquid studies
and thus infer the many-body dynamics. Here we will consider the 
low-energy physics only, i.e. $kd\ll U$. This is important for our
use of (\ref{trep}) for the intratube repulsion that we will return to below.

The effective zero-range coupling
in (\ref{Ueff}) has the property that it (approximately) reproduces the 
$N$-body binding energy of a single chain for large $U$ as discussed 
after (\ref{Ueff}).
However, if one considers
two chains then one realizes that this effective interaction also 
describes the attractive force between two different chains. This follows
readily from the fact that the intertube interactions that are responsible 
for the chain-chain interactions are the same as those that provide the 
intertube attraction in a single chain. If we ignore for a moment any 
intratube repulsion, we can estimate the ground state energy of a 
system with $M$ chains of length $N$ to be simply
\begin{equation}
E_M=\frac{M(M^2-1)}{6}E_{cc}(N)+M E_{cc}(N),
\end{equation}
where $E_{cc}(N)$ is the two-body energy {\it between} 
two {\it different} $N$-chains (consider as effective degrees of 
freedom in the system) calculated using the potential 
$\frac{md^2}{\hbar^2}V_a(x)=-2U_{eff}\delta(\frac{x}{d})$.
The first term is obtained by mapping to the result obtained by 
McGuire \cite{mcguire1964} for $M$ bosonic chains (valid for any $U_{eff}$ and 
thus any $U$) and the second term 
comes from the internal energy of each of the $M$ chains of length $N$. 
The $N$-dependence is given implicitly through $U_{eff}$.

The repulsive intratube interaction has thus far been ignored. We would now
like to include it in a simple way through the use of (\ref{trep}). This 
was derived under the assumptions $U\ll 1$ and $kd\ll U$. The latter is
not a problem, while at first glance $U\ll 1$ is not consistent with the 
fact that we treat the attractive interactions of the chains in a large
$U$ limit. However, we notice that (\ref{trep}) has the correct qualitative
behaviour when $U$ increases; the repulsive potential suppresses transmission
as $T\propto U^{-1}$. We thus assume that we can still use the expression in 
(\ref{trep}) and thus also a delta-function potential with $g=U\lambda^2$.
This is effectively a weak-coupling expression and thus our uncertainty lies
in pushing the attractive part, (\ref{Ueff}), to the weak-coupling limit. 
This has a $20\%$ uncertainty for $U\lesssim 10$ for $N\leq 5$ as noted 
below (\ref{Ueff}). So if we stick to systems with $N\leq 5$ the argument
we have here is only expected to hold to within that uncertainty.

Assuming that we can push the attractive part to weak-coupling and thus 
use a delta-function for both attraction and repulsion, it
is not difficult to take into account through a zero-range interaction term
of the form $\frac{md^2}{\hbar^2}V_r(x)=N U\lambda^2\delta(\frac{x}{d})$. 
The factor of $N$ is obtained
by noting that there will be one repulsive contribution from each tube seperately.
In our effective model of chains, we thus have a competition between two
interactions of opposite sign, $V_a(x)$ and $V_r(x)$. We can use this 
to esimate when the system of chains constitute an attractive or a 
repulsive Bose gas by determining when $V_a(x)=V_r(x)$. Since both
are proportional to $U$, this will give a purely geometric condition. If we
use effective delta-function interactions between chains of length $N$ and define
\begin{equation}\label{func}
F(N,\lambda)=\frac{12}{N(N^2-1)}\left[N H_{N-1}^{(3)}-H_{N-1}^{(2)}\right]-N\lambda^2,
\end{equation}
then $V_a(x)=V_r(x)$ corresponds to $F=0$, while $F>0$ gives an attractive 
and $F<0$ a repulsive chain system. Note that this is not dependent on the number 
of chains, $M$, since we are considering the balance of attraction and repulsion 
by looking at the effective delta-function two-body interactions between two chains
that one would put in the $M$-body Hamiltonian for $M$ chains as the two-body interaction term.
What is important is then the sign of the effective two-body term which is determined
by (\ref{func}) and depends only on the chain length $N$.
We thus see immediately that except for 
very small $\lambda$, the system will be in the repulsive regime for any $N\geq 2$. 
Small $\lambda$ corresponds to a soft transverse confinement that requires 
careful treatment beyond our quasi-1D formalism. We therefore conclude that 
the system of dipolar chains in the intermediate to strong-coupling limit ($U\lesssim 10$) 
with dipoles 
oriented perpendicular to the tubes should behave as a repulsive Tonks-Girardeau
gas. The effective parameters introduced here can be used to estimate ground
state properties using the Lieb-Liniger model \cite{lieb1963a,lieb1963b} and/or
Luttinger liquid formalism can be used to obtain correlation functions etc. 

However, we caution that we mix the limits of strong and weak coupling and 
that carries the uncertainty discussed above. A means of improving this 
is to carry out the derivation of (\ref{trep}) to higher order. That would 
generate a different function of $U$ that could then be inserted in (\ref{func}).
The crucial point is that going to higher order in (\ref{trep}) would give a 
function with higher powers of $U$ in the denominator of (\ref{trep}). 
Consequently, the second term in (\ref{func}) would have a factor $U$ if
we go to second order in (\ref{trep}). Since it corresponds to a 
repulsion it will also not change the sign of the second term in (\ref{func}). 
Since the first term in (\ref{func}) does not carry factors of $U$, this 
means that at higher order the repulsive part will be even stronger and 
we expect this to lead to the same conclusion; $F(N,\lambda)<0$ and 
effectively a repulsive system of chains for larger values of $U$ where
the chains are well bound and thus well-defined.

If we tilt the dipolar orientation away from perpendicular, we expect a 
smaller effective coupling due to the angular factor in (\ref{int}). 
The repulsive part will, however, remain dominant and we expect the 
same conclusion as above to hold for small tilting. If we go to 
the value $\phi_m$ and $\theta=0$ where the intratube repulsion is 
zero, we naturally get an attractively interacting system. However, 
here we need to worry about more complicated complexes than bound states
of chains as we have discussed above (this was first pointed out 
in reference~\cite{wunsch2011}). One could imagine that if we take a 
long chain (large $N$), then it may be possible to attract one more 
particle to bind to this chain around the middle tube. This may 
open the possibility for scattering events between chains where 
this extra particle is exchanged, akin to transfer reactions known 
for instance from nuclear physics. The nature of potentially more 
complicated bound structures
constitute an interesting direction for future work.

\section{Conclusions and outlook}
We have studied dipolar bosonic particles confined to a 
setup consisting of a number of one-dimensional tubes. For the 
case of equidistant adjacent tubes, we calculated the few-body 
bound state structures for up to five particles using both
analytical and numerical tools in the weak and strong binding 
limits. For the case of perpendicularly oriented dipoles, 
we find that chain with one particle in each of a 
number of adjacent tubes are the most stable structures, and that
more complicated bound structures with multiple particles in 
a single tube are unlikely unless one tilts the dipole orientation 
away from the perpendicular direction. The bound states in our
study should be directly observable using experimental techniques
such as lattice shaking \cite{stroh2010}, RF spectroscopy \cite{shin2007,stewart2008}
or {\it in-situ} optical detection \cite{wunsch2011,zinner2011}.

Our results for the weak-coupling limit can be used to calculate the 
low-energy scattering properties of the system and make an effective
zero-range model for bound chains of dipolar bosons that takes
both attractive and repulsive terms into account. This allows us to 
use our knowledge of the few-body physics to address the many-body
physics of these systems which can be mapped onto exactly
solvable model in 1D and Luttinger liquid dynamics. Estimation of the 
magnitudes of the competing attractive and repulsive terms in a system
of chains indicate that for perpendicular orientation the repulsion
is always dominant and a Tonks-Girardeau type of system should emerge. 
This could be changed by tilting the angle of the dipoles with respect to the tubes
which changes the balance between repulsive intratube and attractive intertube
terms. However, the potential presence of bound state
complexes with two or more particles per tube need to be considered in 
this case. The latter will be the focus of future investigations.

In the limit of strong binding we demonstrated that models based on 
replacing the true dipolar potential by a harmonic interaction are 
quite accurate for describing the energetics. This yields exactly solvable
$N$-body models that could for instance be used to study thermodynamic
properties in 1D dipolar systems \cite{jeremy2012a,jeremy2012b}.
Another intriguing possiblity is to study system where one or more of the 
tubes have perpendicular dipoles that point in the opposite direction. 
These flips can be obtained by using AC fields \cite{micheli2007}.
For two dipoles in two adjacent tubes with 
opposite orientation of the dipoles we proved here that in the 
limit of weak dipole moment, there will be no bound two-body 
state. However, this does no in general rule out the possibility
of a bound three-body state, a so-called Borromean state \cite{borromean2012}.
We could thus imagine a regime where three-body states 
are the lowest non-trivial bound states and thus give rise to an effective
gas of trimers. While all of the questions addressed in this paper have 
been concerned with bosonic dipolar particles, we expect that similar 
bound state structures should arise when using fermionic dipoles. The 
fermionic case should correspond closely to studies on one-dimensional 
multi-component systems \cite{guan2010,yin2011} if one maps the component
index onto tube index.


\begin{thebibliography}{11}


\bibitem{ospelkaus2008} Ospelkaus S {\it et al.} 2008 {\it Nature Phys.} {\bf 4} {622}
\bibitem{ni2008} Ni K-K {\it et al.} 2008 {\it Science} {\bf 322} {231}
\bibitem{deiglmayr2008} Deiglmayr J {\it et al.} 2008 {\it Phys. Rev. Lett.} {\bf 101} {133004}
\bibitem{lang2008} Lang F {\it et al.} 2008 {\it Phys. Rev. Lett.} {\bf 101} {133005}
\bibitem{ospelkaus2010} Ospelkaus S {\it et al.} 2010 {\it Science} {\bf 101} {853}
\bibitem{ni2010} Ni K-K {\it et al.} 2010 {\it Nature} {\bf 464} 1324
\bibitem{miranda2011} de Miranda M G H {\it et al.} 2011 {\it Nature Phys.} {\bf 7} {502}
\bibitem{chotia2012} Chotia A {\it et al.} 2012 {\it Phys. Rev. Lett.} {\bf 108} 080405
\bibitem{citro2007} Citro R, Orignac E, De Palo S and Chiofalo M L 2007 {\it Phys. Rev.} A {\bf 75} 051602(R)
\bibitem{citro2008} Citro R, De Palo S,  Orignac E, Pedri P and Chiofalo M L 2008 {\it New. J. Phys.} {\bf 10} 045011
\bibitem{chang2009} Chang C-M {\it et al.} 2009 {\it Phys. Rev.} A {\bf 79} 053630
\bibitem{huang2009} Huang Y-P and Wang D-W 2009 {\it Phys. Rev.} A {\bf 80} 053610
\bibitem{dalmonte2010} Dalmonte M, Pupillo G and Zoller P 2010 {\it Phys. Rev. Lett.} {\bf 105} 140401
\bibitem{kollath2008} Kollath C, Meyer J S and Giamarchi T 2008 {\it Phys. Rev. Lett.} {\bf 100} 130403
\bibitem{fellows2011} Fellows J M and Carr S T 2011 {\it Phys. Rev.} A {\bf 84} 051602(R)
\bibitem{silva2012} De Silva T N 2013 {\it Phys. Lett.} A {\bf 377} 871
\bibitem{dalmonte2011} Dalmonte M, Zoller P and Pupillo G 2011 {\it Phys. Rev. Lett.} {\bf 107} 163202
\bibitem{knap2012} Knap M, Berg E, Ganahl M and Demler E 2012 {\it Phys. Rev.} B {\bf 86} 064501
\bibitem{arguelles2007} Arg{\"u}elles A and Santos L 2007 {\it Phys. Rev.} A {\bf 75} 053613
\bibitem{bauer2012} Bauer M and Parish M M 2012 {\it Phys. Rev. Lett.} {\bf 108} 255302
\bibitem{leche2012} Lecheminant P and Nonne H 2012 {\it Phys. Rev.} B {\bf 85} 195121
\bibitem{tsvelik2012} Tsvelik A M and Kuklov A M 2012 {\it New J. Phys.} {\bf 14} 115033
\bibitem{ruhman2012} Ruhman J, Dalla Torre E G, Huber S D and Altman E 2012 {\it Phys. Rev.} B {\bf 85} 125121
\bibitem{wunsch2011} Wunsch B {\it et al.} 2011 {\it Phys. Rev. Lett.} {\bf 107} 073201
\bibitem{zinner2011} Zinner N T {\it et al.} 2011 {\it Phys. Rev.} A {\bf 84} 063606
\bibitem{santos2010} Klawunn M, Duhme J and Santos L 2010 {\it Phys. Rev.} A {\bf 81} 013604
\bibitem{landau1977} Landau L D and Lifshitz E M 1977 {\it Quantum Mechanics} {(Pergamon Press, Oxford)}
\bibitem{simon1976} Simon B 1976 {\it Ann. Phys. (N.Y.)} {\bf 97} 279
\bibitem{artem2011a} Volosniev A G, Fedorov D V, Jensen A S and Zinner N T 2011 {\it Phys. Rev. Lett.} {\bf 106} 250401 
\bibitem{artem2011b} Volosniev A G, Zinner N T, Fedorov D V, Jensen A S and Wunsch B 2011 {\it J. Phys. B: At. Mol. Opt. Phys.} {\bf 44} 250401
\bibitem{wang2006} Wang D-W, Lukin M D and Demler E 2006 {\it Phys. Rev. Lett.} {\bf 97} 180413
\bibitem{wang2007} Wang D-W 2007 {\it Phys. Rev. Lett.} {\bf 98} 060403
\bibitem{jeremy2010} Armstrong J R, Zinner N T, Fedorov D V and Jensen A S 2010 {\it Europhys. Lett.} {\bf 91} 16001
\bibitem{klawunn2010} Klawunn M, Pikovski A and Santos L 2010 {\it Phys. Rev.} A {\bf 82} 044701
\bibitem{baranov2011} Baranov M A, Micheli A, Ronen S and Zoller P 2011 {\it Phys. Rev.} A {\bf 83} 043602
\bibitem{artem2012} Volosniev A G, Fedorov D V, Jensen A S and Zinner N T 2012 {\it Phys. Rev.} A {\bf 85} 023609
\bibitem{zinner2012} Zinner N T, Wunsch B, Pekker D and Wang D-W 2012 {\it Phys. Rev.} A {\bf 85} 013603
\bibitem{mcguire1964} MacGuire J 1964 {\it J. Math. Phys.} {\bf 5} 622
\bibitem{newton1986} Newton R G 1986 {\it J. Math. Phys.} {\bf 27} 2720
\bibitem{neyenhuis2012} Neyenhuis B {\it et al} 2012 {\it Phys. Rev. Lett.} {\bf 109} 230403
\bibitem{potter2010} Potter A C {\it et al} 2010 {\it Phys. Rev. Lett.} {\bf 105} 220406
\bibitem{pikovski2010} Pikovski A, Klawunn M, Shlyapnikov G V and Santos L 2010 {\it Phys. Rev. Lett.} {\bf 105} 215302
\bibitem{armstrong2011} Armstrong J R, Zinner N T, Fedorov D V and Jensen A S 2013 {\it Few-Body Syst.} {\bf 54} 605
\bibitem{barbara2011} Gapogrosso-Sansone B and Kuklov A 2011 {\it J. Low Temp. Phys.} {\bf 165} 213
\bibitem{armstrong2012} Armstrong J R, Zinner N T, Fedorov D V and Jensen A S 2012 {\it Eur. Phys. J.} D {\bf 66} 85
\bibitem{zinner2012b} Zinner N T, Armstrong J R, Volosniev A G, Fedorov D V and Jensen A S 2012 {\it Few-Body Syst.} {\bf 53} 369
\bibitem{lieb1963a} Lieb E H and Liniger W 1963 {\it Phys. Rev.} {\bf 130} 1605
\bibitem{lieb1963b} Lieb E H 1963 {\it Phys. Rev.} {\bf 130} 1616
\bibitem{suther1971} Sutherland B 1971 {\it J. Math. Phys.} {\bf 12} 246
\bibitem{calogero1971} Calogero F 1971 {\it J. Math. Phys.} {\bf 12} 419
\bibitem{jeremy2011} Armstrong J R, Zinner N T, Fedorov D V and Jensen A S 2011 {\it J. Phys. B} {\bf 44} 055303
\bibitem{jeremy2012} Armstrong J R, Zinner N T, Fedorov D V and Jensen A S 2012 {\it Phys. Scr.} {\bf T151} 014061
\bibitem{artem2011c} Volosniev A G, Armstrong J R, Fedorov D V, Jensen A S and Zinner N T 2013 {\it Few-Body Syst.} {\bf 54} 707
\bibitem{hammer2012} Hammer H-W, Nogga A and Schwenk A 2013 {\it Rev. Mod. Phys.} {\bf 85} 197
\bibitem{frederico2012} Frederico T, Delfino A, Tomio L and Yamashita M T 2012 {\it Prog. Part. Nucl. Phys.} {\bf 67} 939
\bibitem{mcguire1966} MacGuire J 1966 {\it J. Math. Phys.} {\bf 7} 123
\bibitem{tonks1936} Tonks L 1936 {\it Phys. Rev.} {\bf 50} 955
\bibitem{girardeau1960} Girardeau M 1960 {\it J. Math. Phys. (N.Y.)} {\bf 1} 516
\bibitem{deu2010} Deuretzbacher F, Cremon J C and Reimann S M 2010 {\it Phys. Rev.} A {\bf 81} 063616
\bibitem{astra2005} Astrakharchik G E, Boronat J, Casulleras J and Giorgini S 2005 {\it Phys. Rev. Lett.} {\bf 95} 190407
\bibitem{batchelor2005} Batchelor M T, Bortz M, Guan X-W and Oelkers N 2005 {\it J. Stat. Mech.} {\bf 2005} L10001 
\bibitem{calabrese2007} Calabrese P and Caux J-S 2007 {\it Phys. Rev. Lett.} {\bf 98} 150403
\bibitem{cazalilla2011} Cazalilla M A, Citro R, Giamarchi T, Orignac E and Rigol M 2011 {\it Rev. Mod. Phys.} {\bf 83} 1405
\bibitem{haller2009} Haller E {\it et al} 2009 {\it Science} {\bf 325} 1224
\bibitem{haller2010} Haller E {\it et al} 2010 {\it Nature} {\bf 466} 597
\bibitem{stroh2010} Strohmeier N {\it et al.} 2010 {\it Phys. Rev. Lett.} {\bf 104} 080401
\bibitem{shin2007} Shin Y {\it et al.} 2007 {\it Phys. Rev. Lett.} {\bf 99} 090403
\bibitem{stewart2008} Stewart J T, Gaebler J P and Jin D S 2008 {\it Nature} {\bf 454} 744
\bibitem{jeremy2012a} Armstrong J R, Zinner N T, Fedorov D V and Jensen A S 2012 {\it Phys. Rev.} E {\bf 85} 021117
\bibitem{jeremy2012b} Armstrong J R, Zinner N T, Fedorov D V and Jensen A S 2012 {\it Phys. Rev.} E {\bf 86} 021115
\bibitem{micheli2007} Micheli A, Pupillo G, B{\"u}chler H P and Zoller P 2007 {\it Phys. Rev.} A {\bf 76} 043604
\bibitem{borromean2012} Volosniev A G, Fedorov D V, Jensen A S and Zinner N T 2012 {\it Preprint} arXiv:1211.3923
\bibitem{guan2010} Guan L and Chen S 2010 {\it Phys. Rev. Lett.} {\bf 105} 175301
\bibitem{yin2011} Yin X, Guan X-W, Batchelor M T and Chen S 2011 {\it Phys. Rev.} A {\bf 83} 013602

\end{thebibliography}
\end{document}